\documentclass[twocolumn,aps,prl,superscriptaddress,showpacs]{revtex4}
\usepackage{graphicx}
\begin{document}

\newcommand{\hcl}{H_{c1}}
\newcommand{\hcu}{H_{c2}}
\newcommand{\hclo}{H_{c1}^{\rm 1D}}
\newcommand{\hcuo}{H_{c2}^{\rm 1D}}
\newcommand{\s}{\hspace{-1pt}}

\title{Controlling Luttinger Liquid Physics in Spin Ladders under a Magnetic Field}

\author{M. Klanj\v{s}ek}
\affiliation{Grenoble High Magnetic Field Laboratory, CNRS, F-38042 Grenoble Cedex 09, France}

\author{H. Mayaffre}
\affiliation{Laboratoire de Spectrom\'etrie Physique, Universit\'e J. Fourier \& UMR5588 CNRS, F-38402 Saint Martin d'H\`{e}res, France}

\author{C. Berthier}
\affiliation{Grenoble High Magnetic Field Laboratory, CNRS, F-38042 Grenoble Cedex 09, France}

\author{M. Horvati\'c}
\affiliation{Grenoble High Magnetic Field Laboratory, CNRS, F-38042 Grenoble Cedex 09, France}

\author{B. Chiari}
\affiliation{Dipartimento di Chimica, Universit\'a di Perugia, I-06100 Perugia, Italy}

\author{O. Piovesana}
\affiliation{Dipartimento di Chimica, Universit\'a di Perugia, I-06100 Perugia, Italy}

\author{P. Bouillot}
\affiliation{DPMC-MaNEP, University of Geneva, CH-1211 Geneva, Switzerland}

\author{C. Kollath}
\affiliation{Centre de Physique Th\'eorique, Ecole Polytechnique, CNRS, F-91128 Palaiseau Cedex, France}

\author{E. Orignac}
\affiliation{LPENSL CNRS UMR 5672, F-69364 Lyon Cedex 07, France}

\author{R. Citro}
\affiliation{Dipartimento di Fisica ``E. R. Caianiello'' and CNISM, Universit{\`a} di Salerno, I-84100 Salerno, Italy}

\author{T. Giamarchi}
\affiliation{DPMC-MaNEP, University of Geneva, CH-1211 Geneva, Switzerland}

\date{\today}

\begin{abstract}
We present a $^{14}$N nuclear magnetic resonance study of a single crystal of CuBr$_4$(C$_5$H$_{12}$N)$_2$ (BPCB) consisting of weakly coupled spin-$1/2$ Heisenberg antiferromagnetic ladders. Treating ladders in the gapless phase as Luttinger liquids, we are able to fully account for (i) the magnetic field dependence of the nuclear spin-lattice relaxation rate $T_1^{-1}$ at $250$~mK and for (ii) the phase transition to a 3D ordered phase occuring below $110$~mK due to weak interladder exchange coupling. BPCB is thus an excellent model system where the possibility to control Luttinger liquid parameters in a continuous manner is demonstrated and Luttinger liquid model tested in detail over the whole fermion band.
\end{abstract}


\maketitle

Interaction between quantum particles plays a crucial role in one dimension (1D), where its interplay with quantum fluctuations leads to a state described as a Luttinger liquid (LL) \cite{Giamarchi_2004}. Low-energy physics of the LL is fully characterized by two interaction dependent LL parameters: the velocity of excitations, $u$, and the dimensionless exponent $K$. Correlation functions decay as power laws, with exponents which are simple functions of $K$. The LL model has been shown to apply to a growing number of 1D systems, such as organic conductors \cite{Schwartz_1998}, quantum wires \cite{Auslaender_2002}, carbon nanotubes \cite{Ishii_2003}, edge states of quantum Hall effect \cite{Grayson_1998}, ultra cold atoms \cite{Bloch_2008}, and antiferromagnetic (AFM) spin chain \cite{Lake_2005} or spin ladder systems \cite{Dagotto_1998}. Several characteristic features of the LL model have been observed in these systems, such as the power law behavior of some correlation or spectral functions. However, since the details of the interaction are rarely known, only a theoretical estimate of the power law exponents is usually possible. A precise quantitative check of the LL model is thus still missing.

\vspace{-0.1cm}

The obstacle can be overcome in spin ladder systems. Namely, spin-$1/2$ AFM ladder in an external magnetic field $H$ maps essentially onto a 1D system of interacting spinless fermions \cite{Chitra_1997, Giamarchi_1999, Hikihara_2001, Giamarchi_2004}, where $H$ acts as a chemical potential. The interaction term in the fermion picture is {\em uniquely} determined by the exchange coupling constants, which can be experimentally extracted [$J_\perp$ on the rungs and $J_\parallel$ on the legs of the ladder; see Fig.~1(a)], and by $H$, which controls the filling of the fermion band. By increasing $H$ the spin gap between the singlet ground state and the lowest triplet ($S_z=-1$) excited states of the spin ladder decreases. It closes at $\hcl$, where the ladder enters the gapless phase corresponding to the partially filled fermion band. At $\hcu$ the ladder gets fully polarized and the gap reopens. Once $J_\perp$ and $J_\parallel$ are known the LL parameters can be obtained numerically for arbitrary $H$. The associated LL prediction can then be checked {\em quantitatively} over the {\em whole} fermion band, which extends between the critical fields $\hcl$ and $\hcu$.

\vspace{-0.1cm}

\begin{figure}[b]
\includegraphics[width=0.35\textwidth]{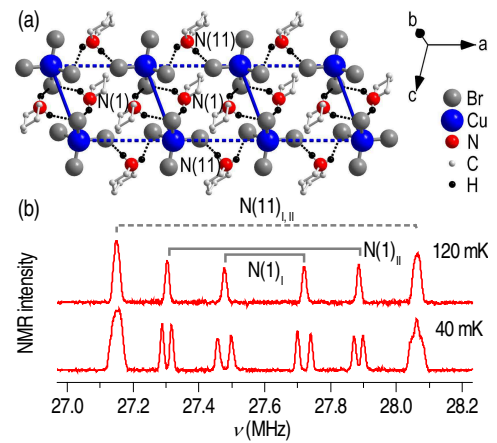}
\caption{
(color online). (a) A ladder formed by the supposed exchange interactions, $J_{ij} \mathbf{S}_i\cdot\mathbf{S}_j$, between $S=1/2$ spins of Cu$^{2+}$ ions in the crystal structure of BPCB. Solid and dashed thick blue lines stand for $J_\perp$ and $J_\parallel$, respectively. The 10 protons attached to the C atoms are not shown. (b) $^{14}$N NMR spectra at $120$ and $40$~mK recorded at $H=9.0$~T. 
}
\end{figure}

Suitable spin ladder systems are rare, either because of unattainable critical fields or because of the presence of anisotropic interactions, such as Dzyaloshinski-Moriya (DM) interaction \cite{Clemancey_2006}, as is the case in extensively studied Cu$_2$(C$_5$H$_{12}$N$_2$)$_2$Cl$_4$ \cite{Chaboussant_1998}. Recently, CuBr$_4$(C$_5$H$_{12}$N)$_2$ (BPCB) \cite{Patyal_1990} was identified as a good candidate. Namely, low-temperature magnetization data were well described by the $XXZ$ chain model \cite{Watson_2001} in the strong-coupling limit ($J_\perp\gg J_\parallel$) of a ladder \cite{Chaboussant_1998, Mila_1998, Giamarchi_1999}. In addition, the results of thermal expansion and magnetostriction experiments were explained within the free fermion model and in more detail with quantum Monte Carlo calculations for spin ladders \cite{Lorenz_2008, Anfuso_2008}. In this Letter we present a detailed $^{14}$N nuclear magnetic resonance (NMR) study of BPCB revealing for the first time the occurence of the field induced 3D magnetic order below $110$~mK. We show that the LL model completely accounts for the experimental behavior of BPCB in the gapless phase. The determined phase diagram and field variation of the order parameter are perfectly described in the framework of weakly coupled LLs. The importance of these results is twofold: (i) they show that the whole physics of coupled spin ladders can be captured in a {\em single} theory based on LLs, and (ii) they provide the first quantitative check of the LL model.

In $^{14}$N NMR experiments we used a single crystal of BPCB with dimensions 3.5$\times$2$\times$4~mm$^3$. As shown in Fig.~1(a), the pairs of spin-$1/2$ Cu$^{2+}$ ions (spin dimers) in the crystal structure of BPCB are stacked along the crystallographic $a$ axis to form ladders \cite{Patyal_1990}. Each unit cell contains two rungs of two different ladders (denoted by I and II), which are crystallographically equivalent, but become physically inequivalent when the external magnetic field $\mathbf{H}$ is applied in an arbitrary direction. There are only two crystallographically inequivalent nitrogen (N) sites per ladder in a unit cell: N(1) is located close to the rung and N(11) close to the leg of the ladder [Fig.~1(a)]. Since $^{14}$N has spin $I=1$ and thus a quadrupole moment, each site gives rise to a pair of $^{14}$N NMR lines (doublet), split by the quadrupole coupling with the local environment. Therefore, $^{14}$N NMR spectrum [upper spectrum in Fig.~1(b)] consists of four doublets, for two sites in each of the two ladders. We oriented the sample so that the external magnetic field lay in the $a^*b$ plane ($a^*\perp b,c$), at an angle of $9^\circ$ to the $b$ axis. In this orientation, the N(1) doublets are well resolved, while the N(11) doublets overlap [Fig.~1(b)]. All the measurements presented here were performed on the N(1)$_{\rm I}$ lines.

The magnetization of the Cu$^{2+}$ ions is detected via the hyperfine coupling to the $^{14}$N nuclei. It is easily reconstructed due to the inversion center in the middle of each rung \cite{Patyal_1990}. Any uniform magnetization of Cu$^{2+}$ ions {\em shifts} the average position of the NMR quadrupole doublet with respect to the Larmor frequency. Any staggered (i.e., AFM) magnetization of Cu$^{2+}$ ions, however, breaks the inversion symmetry between the pairs of equivalent N sites. This doubles the number of inequivalent N sites and each NMR line {\em splits} symmetrically in two lines. The associated hyperfine shift and splitting are proportional to the uniform and staggered magnetization of Cu$^{2+}$ ions, respectively. Fig.~1(b) illustrates the emergence of the staggered magnetization in BPCB at $H=9.0$~T on cooling from $120$ to $40$~mK. Its origin is discussed later.

\begin{figure}[b]
\includegraphics[width=0.35\textwidth]{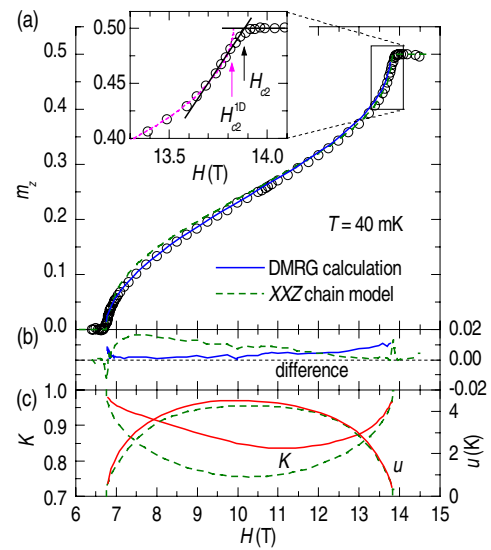}
\caption{
(color online). (a) Magnetic field dependence of the longitudinal uniform magnetization $m_z$ per Cu$^{2+}$ ion at $40$~mK ({\large $\circ$}). Inset shows linear $m_z(H)$ dependence very close to $\hcu$ (solid line) and the determination of $\hcuo$ (dashed line). The data are compared to the result of the DMRG calculation for $J_\perp/J_\parallel =3.6$ (solid line) and to the prediction of the $XXZ$ chain model (dashed line), both at $T=0$. (b) Difference between each prediction and the experimental data. (c) Variation of the LL parameters $K(H)$ and $u(H)$ (in kelvin units) over the fermion band as calculated for $J_\perp/J_\parallel =3.6$ (solid line) and for the $XXZ$ chain model (dashed line).
}
\end{figure}

Next we check whether our longitudinal (parallel to $\mathbf{H}$) uniform magnetization $m_z(H)$ data are well described in the $XXZ$ chain model. Fig.~2(a) shows $m_z(H)$ (per spin) measured at $40$~mK via the hyperfine shift of N(1)$_{\rm I}$ NMR lines, from which we extract the values of the critical fields: $\hcl=6.703\pm 0.008$~T and $\hcu =13.888\pm 0.006$~T. Instead of the square-root singularities expected for a perfect 1D system, very close to the critical fields we observe a linear $m_z(H)$ dependence [Fig.~2(a) inset], a signature of weak inter-ladder (3D) exchange coupling \cite{Giamarchi_1999}. As this 3D regime is very narrow, by fitting $m_z(H)$ data with a phenomenological function having square-root singularities at $m_z=0$ and $0.5$ [Fig.~2(a) inset], we can estimate the values of the corrected, ``1D'' critical fields pertaining to an {\em isolated} ladder: $\hclo=6.763$~T and $\hcuo=13.828$~T. With $g=2.176$ corresponding to our sample orientation \cite{Patyal_1990} this yields \cite{Reigrotzki_1994} $J_\perp =12.9$~K and $J_\parallel =3.6$~K, hence $J_\perp/J_\parallel =3.6$, in agreement with recent determination \cite{Lorenz_2008, Anfuso_2008}. With these values we perform the density matrix renormalization group (DMRG) calculation of $m_z(H)$ at $T=0$ for a single ladder and find an excellent agreement with our experimental data. As shown in Figs.~2(a,b), the slight asymmetry of the curve about its mid-point at $m_z=0.25$ is nicely reproduced since we take into account also the upper two triplet states of each dimer ($S_z=0,1$). These are neglected in the strong-coupling treatment ($J_\perp\gg J_\parallel$) leading to the $XXZ$ chain model \cite{Giamarchi_1999}. The corresponding curve is symmetric about its mid-point, in worse agreement with our experimental data. However, $m_z(H)$ as a thermodynamic quantity is not very sensitive to the model. Once the coupling ratio $J_\perp/J_\parallel$ is fixed, the variation of the LL parameters with $H$ is completely determined and the LL theory is left without any adjustable parameter. For an isolated ladder with $J_\perp/J_\parallel =3.6$ we numerically calculate $K(H)$ and $u(H)$ (in kelvin units), combining DMRG method with bosonization as in Ref.~\cite{Hikihara_2001}. The result is displayed in Fig.~2(c) together with the corresponding result in the $XXZ$ chain model (from Ref.~\cite{Giamarchi_1999}). Close to $\hclo$ ($\hcuo$) the LL exponent approaches the value $K=1$ of noninteracting fermion system indicating a nearly empty (full) fermion band.

The LL behavior is tested via the dynamical spin-spin correlation functions, which are experimentally accessible through NMR observables. We focus on three such observables, starting with the nuclear spin-lattice relaxation rate $T_1^{-1}$. At low temperatures $T_1^{-1}$ probes exclusively Cu$^{2+}$ spin dynamics, namely its low-energy ($\omega\rightarrow 0$) part \cite{Moriya_1956} corresponding to the long time behavior of the local spin-spin correlation functions. For the above determined range of $K(H)$, by far the biggest contribution to $T_1^{-1}$ at low temperature comes from the transverse (perpendicular to $\mathbf{H}$) staggered correlation \cite{Giamarchi_1999}, and we find it to be \cite{Giamarchi_2004, Giamarchi_1999}:
\begin{equation}
T_1^{-1} = \frac{\hbar\gamma^2 A_\perp^2 A_0^x}{k_B u}
\cos\s\s\left(\s\frac{\pi}{4K}\s\right)
B\s\s\left(\s\s \frac{1}{4K}, 1\s\s-\s\s\frac{1}{2K} \s\s\right)\s\s
\left(\s\s \frac{2\pi T}{u} \s\right)^{\s\s\frac{1}{2K}-1},
\label{eqT1}
\end{equation}
where $A_0^x$ is the amplitude of the correlation function, $\gamma/(2\pi)=3.076$~MHz/T is $^{14}$N nuclear gyromagnetic ratio, $A_\perp$ the transverse hyperfine coupling constant, and $B(x,y)=\Gamma(x)\Gamma(y)/\Gamma(x+y)$. Fig.~3(a) shows $^{14}$N $T_1^{-1}(H)$ dependence measured in the gapless phase of BPCB at $250$~mK, well above the 3D ordering temperature. Its concave shape reflects the increased 1D fermion density of states close to the critical fields. Using the previously calculated $K(H)$, $u(H)$ [from Fig.~2(c)] and $A_0^x(H)$ (calculated along with $K$ and $u$) for $J_\perp/J_\parallel =3.6$, $T_1^{-1}(H)$ from Eq.~(\ref{eqT1}) is compared to the data by adjusting a {\em single} scaling factor $A_\perp^2$ [Fig.~3(a)]. Very good agreement over the whole field range provides a remarkable confirmation of the LL model. Moreover, the utilized value $A_\perp =570$~G agrees with that obtained from direct $^{14}$N NMR determination \cite{Mayaffre_2007}. In contrast, the curve obtained in the $XXZ$ chain model (with the same $A_\perp$) fails to reproduce the biased shape of $T_1^{-1}(H)$ [Fig.~3(a)]. This demonstrates the sensitivity of the observable $T_1^{-1}(H)$ to the applied set of LL parameters $K(H)$, $u(H)$ and, consequently, to the coupling ratio $J_\perp/J_\parallel$.

\begin{figure}
\includegraphics[width=0.32\textwidth]{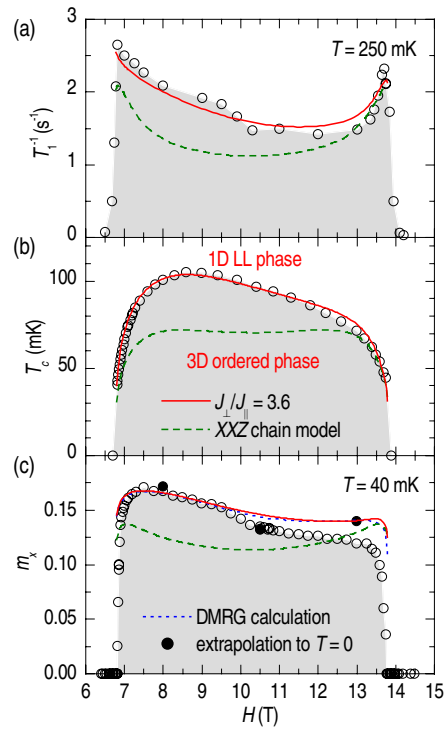}
\caption{
(color online). Magnetic field dependence of (a) $^{14}$N $T_1^{-1}$ at $250$~mK ({\large $\circ$}), in the 1D LL phase, (b) the temperature of the transition between the 3D ordered phase and the 1D LL phase ({\large $\circ$}), (c) the transverse staggered magnetization $m_x$ per Cu$^{2+}$ ion at $40$~mK ({\large $\circ$}) and its extrapolation to zero temperature ({\large $\bullet$}), all measured on N(1)$_1$ site as described in the text. Each data set is compared to the corresponding prediction of the LL model based on $K(H)$ and $u(H)$ [from Fig.~2(c)] for $J_\perp/J_\parallel =3.6$ (solid red line) and to the prediction of the simplified $XXZ$ chain model (dashed green line). The {\em only} adjustable parameter in each case is an overall scaling factor. Dotted blue line in (c) is the result of the DMRG calculation.
}
\end{figure}

The (dominant) transverse staggered spin-spin correlation function diverges with decreasing temperature, which leads at low enough temperature to a 3D ordering due to weak interladder exchange coupling \cite{Giamarchi_1999}. The associated order parameter is a transverse staggered magnetization $m_x$ (per spin) \cite{footnote_1}, measured via the hyperfine splitting of the N(1)$_{\rm I}$ NMR lines. To map the boundary between the 3D ordered phase and the 1D LL phase, we determine the temperatures $T_c(H)$, at which this splitting vanishes [Fig.~3(b)]. Since the whole phase boundary lies below $110$~mK ($\ll J_\parallel =3.6$~K), we expect that the interladder coupling (i.e., the coupling between the LLs) is small and treat it in the mean-field approximation leading to \cite{Giamarchi_1999}:
\begin{equation}
T_c = \frac{u}{2\pi} \left[
\sin\s\left(\s\frac{\pi}{4K}\s\right) B^2\s\s\left(\s\s \frac{1}{8K}, 1\s\s-\s\s\frac{1}{4K} \s\s\right)
\frac{z J' A_0^x}{2u}
\right]^\frac{2K}{4K-1}.
\label{eqTc}
\end{equation}
Here $J'$ is the exchange coupling between the Cu$^{2+}$ ions in neighboring ladders and $z$ the coordination number. Although exchange paths between the ladders have not been identified yet, the structure of BPCB suggests that $z=4$. Using $K(H)$, $u(H)$ and $A_0^x(H)$ for $J_\perp/J_\parallel =3.6$, we compare $T_c(H)$ from Eq.~(\ref{eqTc}) to the experimental data by adjusting a {\em single} scaling parameter $J'$ [Fig.~3(b)]. For $J'=20$~mK \cite{footnote_2} we obtain an excellent agreement, with faithfully reproduced biased shape. The prediction of the $XXZ$ chain model (with the same $J'$) is considerably different in shape, showing that $T_c(H)$ is another observable very sensitive to the applied set of LL parameters.

The field dependence of the order parameter $m_x(H)$ is measured via the hyperfine splitting of the N(1)$_{\rm I}$ NMR lines at $40$~mK [open circles in Fig.~3(c)]. The data are compared to the $T=0$ prediction, which we obtain in the mean-field treatment based on weakly coupled LLs:
\vspace{-0.1cm}
\begin{equation}
m_x = F(K) \sqrt{A_0^x} \left( \frac{\pi z J' A_0^x}{2u} \right)^{1/(8K-2)},
\label{eqmx}
\end{equation}
where, using the results from Ref. \cite{Lukyanov_1997},
\begin{displaymath}
F(K) =
\Biggl\{
\frac{
\frac{\pi^2}{\sin(\pi/(8K-1))} \frac{8K}{8K-1}
\bigl[ \frac{\Gamma(1-(1/8K))}{\Gamma(1/8K)} \bigr]^\frac{8K}{8K-1}
}{
\Gamma\bigl(\frac{4K}{8K-1}\bigr)^2 \Gamma\bigl(\frac{16K-3}{16K-2}\bigr)^2
}
\Biggr\}^\frac{8K-1}{8K-2}.
\end{displaymath}
Taking again $K(H)$, $u(H)$, $A_0^x(H)$ for $J_\perp/J_\parallel =3.6$ and $J'=20$~mK, the analytical $m_x(H)$ dependence defined by Eq.~(\ref{eqmx}) is {\em fixed}. As the proportionality constant between $m_x$ and the measured hyperfine splitting is not known, in Fig.~3(c) we scale the data in vertical direction to match the analytical curve at the maximum near $7.5$~T. The trend for an overall field dependence is nicely reproduced except for some discrepancy between $10$~T and $13$~T. However, the theoretical $T=0$ prediction is here compared to the data recorded at $40$~mK, which may not be fully saturated to the zero temperature limit. Indeed, an extrapolation to $T=0$ [solid circles in Fig.~3(c)] from the temperature dependence measured above $40$~mK indicates that the agreement is better than what appears to be. In addition, we calculate $m_x(H)$ by DMRG, treating again the interladder coupling with $J'=20$~mK in the mean-field approximation. As shown in Fig.~3(c), the result is hardly distinguishable from the analytical curve.

The field induced 3D magnetic ordering in the gapless phase of BPCB below $110$~mK is thus perfectly described in the framework of weakly coupled 1D systems, treated as LLs. Both the phase boundary $T_c(H)$ and the field variation of the order parameter $m_x(H)$ exhibit a biased shape. This contrasts with the dome shape of both observables from essentially 3D spin dimer systems, such as TlCuCl$_3$ \cite{Nohadani_2004} or BaCuSi$_2$O$_6$ \cite{Jaime_2004}. Moreover, $m_x(H)$ in BPCB is {\em not} proportional to $T_c(H)$, contrary to what is expected for the standard mean-field description of a phase transition in isotropic 3D systems. This is another signature of the underlying 1D physics. However, at low enough temperature, close to the critical fields the LL description in the gapless phase collapses as the system restores its full 3D character \cite{Giamarchi_1999}. In this region the low-temperature 3D magnetic ordered state is predicted to be a Bose-Einstein condensate \cite{Giamarchi_1999, Nikuni_2000} leading to the linear $m_z(H)$ dependence \cite{Giamarchi_1999}, which is indeed observed in our experiment.

In summary, we have investigated the experimental behavior of BPCB in the gapless phase, including 3D magnetic ordering occuring below $110$~mK. All our experimental results are perfectly described in a {\em single} theory based on LLs, with {\em no} adjustment or fitting of the LL parameters. Magnetic field variation of $T_1^{-1}$, of the transition temperature and of the order parameter (at $T=0$), which are shown to depend substantially on the coupling ratio $J_\perp/J_\parallel$, provide sensitive probes for the behavior of the dominant spin-spin correlation function and its role in low temperature 3D magnetic ordering. A {\em single} set of LL parameters is shown to control {\em all} three observables, which is an essential feature of the LL model.

We are grateful to A.~Furusaki and T.~Hikihara for providing us their unpublished results. This work was supported in part by French ANR grant 06-BLAN-0111, by the European Community contract RITA-CT-2003-505474, by the Swiss NSF, under MaNEP and Division II, and by the RTRA network ``Triangle de la Physique''.


\begin{thebibliography}{22}

\bibitem{Giamarchi_2004}
T.~Giamarchi, {\it Quantum Physics in One Dimension} (Oxford Univ. Press, Oxford, 2004).

\bibitem{Schwartz_1998}
A.~Schwartz {\em et al.}, Phys. Rev. B {\bf 58}, 1261 (1998).

\bibitem{Auslaender_2002}
O.~M.~Auslaender {\em et al.}, Science {\bf 295}, 825 (2002).

\bibitem{Ishii_2003}
H.~Ishii {\em et al.}, Nature (London) {\bf 426}, 540 (2003).

\bibitem{Grayson_1998}
M.~Grayson {\em et al.}, Phys. Rev. Lett. {\bf 80}, 1062 (1998).

\bibitem{Bloch_2008}
I.~Bloch, J.~Dalibard, and W.~Zwerger, Rev. Mod. Phys. {\bf 80}, 885 (2008).

\bibitem{Lake_2005}
B.~Lake {\em et al.}, Nature Mater. {\bf 4}, 329 (2005).

\bibitem{Dagotto_1998}
E.~Dagotto, Rep. Prog. Phys. {\bf 62}, 1525 (1999).

\bibitem{Chitra_1997}
R.~Chitra and T.~Giamarchi, Phys. Rev. B {\bf 55}, 5816 (1997).

\bibitem{Giamarchi_1999}
T.~Giamarchi and A.~M.~Tsvelik, Phys. Rev. B {\bf 59}, 11398 (1999).

\bibitem{Hikihara_2001}
T.~Hikihara and A.~Furusaki, Phys. Rev. B {\bf 63}, 134438 (2001).

\bibitem{Clemancey_2006}
M.~Cl\'emancey {\em et al.}, Phys. Rev. Lett. {\bf 97}, 167204 (2006).

\bibitem{Chaboussant_1998}
G.~Chaboussant {\em et al.}, Eur. Phys. J. B {\bf 6}, 167 (1998).

\bibitem{Patyal_1990}
B.~R.~Patyal, B.~L.~Scott, and R.~D.~Willett, Phys. Rev. B {\bf 41}, 1657 (1990).

\bibitem{Watson_2001}
B.~C.~Watson {\em et al.}, Phys. Rev. Lett. {\bf 86}, 5168 (2001).

\bibitem{Mila_1998}
F.~Mila, Eur. Phys. J. B {\bf 6}, 201 (1998).

\bibitem{Lorenz_2008}
T.~Lorenz {\em et al.}, Phys. Rev. Lett. {\bf 100}, 067208 (2008).

\bibitem{Anfuso_2008}
F. Anfuso {\em et al.}, Phys. Rev. B {\bf 77}, 235113 (2008).

\bibitem{Reigrotzki_1994}
M.~Reigrotzki, H.~Tsunetsugu, and T.~M.~Rice, J. Phys.: Condens. Matter {\bf 6}, 9235 (1994).

\bibitem{Moriya_1956}
T.~Moriya, Prog. Theor. Phys. {\bf 16}, 23 (1956).

\bibitem{Mayaffre_2007}
H.~Mayaffre {\em et al.} (unpublished).

\bibitem{footnote_1}
Because of the absence of a Dzyaloshinski-Moriya interaction for a spin dimer, $m_x$ only appears in the ordered phase (see Ref.~\cite{Clemancey_2006}).

\bibitem{footnote_2}
We remark that $J'$ is an average value of inter-ladder couplings, which may be compensating each other. The mean-field approximation probably slightly underestimates the absolute value of $J'$.

\bibitem{Lukyanov_1997}
S.~Lukyanov and A.~Zamolodchikov, Nucl. Phys. B {\bf 493}, 571 (1997).

\bibitem{Nohadani_2004}
O.~Nohadani {\em et al.}, Phys. Rev. B {\bf 69}, 220402(R) (2004).

\bibitem{Jaime_2004}
M.~Jaime {\em et al.}, Phys. Rev. Lett. {\bf 93}, 087203 (2004).

\bibitem{Nikuni_2000}
T.~Nikuni {\em et al.}, Phys. Rev. Lett. {\bf 84}, 5868 (2000).

\end{thebibliography}
\end{document}